
\magnification=1200
\hoffset=-.1in
\voffset=-.2in

\vsize=7.5in
\hsize=5.6in
\tolerance 10000

\baselineskip 12pt plus 1pt minus 1pt
\pageno=0

\centerline{{\bf GAUGE INVARIANT FORMULATIONS OF LINEAL
GRAVITY}\footnote{*}{This
work is supported in part by funds
provided by the U. S. Department of Energy (D.O.E.) under contract
\#DE-AC02-76ER03069, and (for D.C.) by the Swiss National Science Foundation.}}
\vskip 24pt
\centerline{D. Cangemi and R. Jackiw}
\vskip 12pt
\centerline{\it Center for Theoretical Physics}
\centerline{\it Laboratory for Nuclear Science}
\centerline{\it and Department of Physics}
\centerline{\it Massachusetts Institute of Technology}
\centerline{\it Cambridge, Massachusetts\ \ 02139\ \ \ U.S.A.}
\vskip 1.5in
\centerline{Submitted to: {\it Physical Review Letters\/}}
\vfill
\noindent hepth@xxx/9203056 \hfill
\centerline{ Typeset in $\TeX$ by Roger L. Gilson}
\vskip -12pt
\noindent CTP\#2085\hfill March 1992
\eject
\baselineskip 24pt plus 2pt minus 2pt
\centerline{\bf ABSTRACT}
\medskip
It is shown that the currently studied ``string-inspired'' model for gravity
on a line can be formulated as a gauge invariant theory based on the Poincar\'e
group with central extension --- a formulation that complements and simplifies
H.~Verlinde's construction based on the unextended Poincar\'e group.
\vfill
\eject
Lineal gravities, {\it i.e.\/} Einstein-type theories in $(1+1)$-dimensional
space-time, provide a setting for studying un-understood issues of
gravitation.  The simplifications achieved by the drastic dimensional
reduction are not devoid of interest, provided the dynamical equations are not
based on the Einstein tensor $R_{\mu\nu} - {1\over 2} g_{\mu\nu}
R$, which vanishes identically in two dimensions.  Several
years ago, a class of theories based on the Riemann
scalar $R$ was proposed, but even the simplest of these,$^1$ in which scalar
curvature is equated to a cosmological constant $\Lambda$,
$$R - \Lambda = 0 \eqno(1)$$
requires an additional, non-geometrical field in an action formulation: Eq.~(1)
follows from the action
$$I_1 = \int d^2x \sqrt{-g}\,\eta (R-\Lambda) \eqno(2)$$
where $\eta$ is an invariant world-scalar, which acts as a Lagrange multiplier
enforcing (1).  Of course, once the additional scalar field has been
introduced, one may consider various generalizations and modifications of (1),
(2) with alternative dynamics for $R$ and $\eta$.$^2$

The model (1), (2) has two distinctive features.  It can be obtained by
dimensional reduction from Einstein theory in three space-time dimensions.$^1$
Moreover, what is of particular interest for us here, it possesses a gauge
theoretical formulation, given by Isler and Trugenberger as well as by
Chamseddine
and Wyler.$^3$  To this end one uses the de~Sitter or anti-de~Sitter group
with Lorentz generator $J$ and
translation generators $P_a$ satisfying the $SO(2,1)$ algebra (for
$\Lambda\not=0$).
$$\left[ P_a, J\right] = \epsilon_a{}^b P_b\ \ ,\qquad
\left[ P_a, P_b\right] = - {1\over 2}\Lambda \epsilon_{ab} J \eqno(3)$$
[The tangent space indices ($a,b,\ldots$) are
raised and lowered with the flat space metric tensor $h_{ab}=\hbox{diag}(1,-1)$
and $\epsilon^{01}=1$.]   In the usual way, the gauge connection 1-form $A$ is
expanded in terms of the generators,
$$A = e^a P_a + \omega J\eqno(4)$$
where $e^a_\mu$ is the {\it Zweibein\/} and $\omega_\mu$ is the
spin connection.  The curvature 2-form
$$F = dA + A^2 \eqno(5)$$
becomes
$$F = f^a P_a + fJ = (De)^a P_a + \left( d\omega - {1\over 4}\Lambda e^a
\epsilon_{ab} e^b\right) J \eqno(6)$$
$$(De)^a \equiv de^a + \epsilon^a{}_b \omega e^b \eqno(7)$$
The three field strengths $F^A=(f^a,f)$ transform covariantly according to the
three-dimensional adjoint representation.  Therefore
the Lagrange density
$$\eqalign{{\cal L}'_1 &= \sum\limits^2_{A=0} \eta_A F^A = \eta_a (De)^a +
\eta_2 \left( d\omega -
{1\over 4}\Lambda e^a \epsilon_{ab} e^b\right) \cr
\eta_A &= \left( \eta_a,\eta_2\right)\cr}\eqno(8)$$
is gauge invariant when the
Lagrangian multiplier triplet $\eta_A$ is taken to
transform by the coadjoint representation.  The
equation obtained by varying $\eta_a$ allows evaluating the spin
connection in terms of the {\it Zweibein\/},
$$\omega = e^a \left( h_{ab} \epsilon^{\mu\nu} \partial_\mu
e^b_\nu\right)\big/ \det e \eqno(9)$$
and the equation that follows upon variation of $\eta_2$ regains (1).
Finally it is noted that a non-degenerate Killing metric is available because
the relevant group is semi-simple for $\Lambda\not=0$.

Recently, Verlinde$^4$ as well as Callan, Giddings, Harvey and Strominger$^5$
have introduced a similar model, which is ``string-inspired.''  The action
$$I_2 = \int d^2x\,\sqrt{-g} \left( \eta R - \Lambda\right) \eqno(10)$$
differs from (2) in that the Lagrange multiplier is absent from the
cosmological constant.$^6$  The equation
of motion from varying $\eta$
$$R = 0 \eqno(11)$$
shows that the metric is flat, $g_{\mu\nu} = h_{\mu\nu}$, while varying
$g_{\mu\nu}$ gives, with the help of (11)
$$\partial_\mu \partial_\nu \eta = {1\over 2}\Lambda h_{\mu\nu}  \eqno(12)$$
Thus
$$-2\eta = M - \Lambda\left( x^+ - x^+_0\right) \left( x^- - x^-_0\right) \ \
,\qquad x^\pm = {1\over \sqrt{2}}(t\pm x)\ \ ,\eqno(13)$$
with $x^\pm_0$ and $M$ being integration constants.  Interest in the model
derives from the above ``black-hole'' solution$^7$ with mass $M$ [in terms of
the ``physical'' metric $ g_{\mu\nu}/(-2\eta)$] whose quantum mechanical
analysis may shed light on various quantum gravity puzzles.$^{4,\,5,\,8}$

Here we address the problem of a gauge theoretical formulation for $I_2$.  A
discussion has already been given by Verlinde,$^4$ based on the
non semi-simple Poincar\'e group, with the algebra
$$\left[ P_a,J\right] = \epsilon_a{}^b P_b \ \ ,\qquad
\left[ P_a, P_b\right] = 0 \eqno(14)$$
which is the $\Lambda\to0$ contraction
of (3).  However, there are various unexpected
features in his formulation: the transformation law for the Lagrange
multipliers
is an unfamiliar affine expression; the Lagrange density is not invariant but
changes by a total derivative --- see below.  After reviewing Verlinde's
approach, we show that an {\it invariant\/} Lagrange density with a
conventional coadjoint transformation for the Lagrange multipliers can be
given, provided one uses a {\it centrally extended\/} Poincar\'e algebra,
which is an unconventional contraction of (3).

Following Verlinde, the connection and curvature are defined as in (4)--(7),
but owing to the vanishing of the momentum commutator, the curvature
becomes the $\Lambda=0$ limit of (6).
$$F = f^a P_a + fJ = (De)^a P_a + d\omega J\eqno(15)$$
Infinitesimal gauge transformation rules
$$\delta A = d\Theta + \left[ A,\Theta\right] \eqno(16)$$
with the gauge generator $\Theta$
$$\Theta = \theta^a P_a + \alpha J \eqno(17)$$
can be deduced from (14), (16) to be
$$\eqalign{\delta e^a &= - \alpha
\epsilon^a{}_b e^b + \epsilon^a{}_b  \theta^b\omega + d\theta^a\cr
\delta\omega &= d\alpha\cr} \eqno(18)$$
In finite form they read
$$\eqalign{e^a \to \bar{e}^a &= \left( {\cal M}^{-1}\right)^a_{\ b}
\left( e^b +  \epsilon^b{}_c \theta^c\omega + d\theta^b\right) \cr
\omega \to \bar{\omega} &= \omega+ d\alpha\cr}\eqno(19)$$
where ${\cal M}$ is a finite Lorentz transformation.
$${\cal M}^a{}_b = \delta^a{}_b \cosh \alpha + \epsilon^a{}_b \sinh \alpha
\eqno(20)$$
The curvature triplet $F^A$ transforms according to the three-dimensional
adjoint representation of the Poincar\'e group, {\it viz.\/} as in (19) but
without the differentials.
$$\eqalign{f^a \to \bar{f}^a &= \left( {\cal M}^{-1}\right)^a_{\ b} \left( f^b
+ \epsilon^b{}_c \theta^c f\right) \cr
\bar{f} \to \bar{f} &= f \cr}\eqno(21\hbox{a})$$
or
$$\eqalign{ F^A\to \bar{F}^A
&= \sum\limits^2_{B=0} \left( U^{-1}\right)^A_{\  B} F^B
\cr
U &= \left( \matrix{ {\cal M}^a{}_b & - \epsilon^a{}_c \theta^c
\cr\noalign{\vskip 0.2cm} 0 & 1 \cr}\right) \cr}\eqno(21\hbox{b})$$
The Lagrange density is now taken as
$$\eqalign{ {\cal L}'_2 &= \sum^2_{A=0} \eta_A F^A + {1\over 2}\Lambda\, e^a
\epsilon_{ab} e^b \cr
&= \eta_a (De)^a + \eta_2\,d\omega + {1\over 2}\Lambda\, e^a \epsilon_{ab}
e^b\cr}\eqno(22)$$
The equations of motion that follow from ${\cal L}'_2$ are equivalent to
(11) -- (13) (with $\eta_2=-2\eta$).  However,
the transformation properties of ${\cal L}'_2$ under Poincar\'e
gauge transformations are obscure.  Following Verlinde, we can check that the
following infinitesimal rules for $\delta\eta_A$
$$\delta \eta_a = \eta_b\epsilon^b{}_a\alpha - \Lambda \epsilon_a{}_b
\theta^b\ \ ,\qquad \delta \eta_2 = -\eta_a \epsilon^a{}_b \theta^b
\eqno(23)$$
together with (18) change ${\cal L}'_2$ by a total derivative.  But the affine
shift in $\delta\eta_a$, proportional to $\Lambda$, is unfamiliar.  In the
finite version of (23)
$$\eqalign{\eta_a \to \bar{\eta}_a &= \left( \eta_b - \Lambda \epsilon_{bc}
\theta^c\right) {\cal M}^b{}_a\cr
\eta_2 \to \bar{\eta}_2 &= \eta_2 - \eta_a \epsilon^a{}_b \theta^b - {1\over
2}\Lambda\theta^2 \cr}\eqno(24)$$
the homogenous part is a coadjoint transformation ($\bar{\eta} = \eta
U)$, while
the shift, proportional to $\Lambda$, is needed to compensate for the gauge
non-invariance of the cosmological constant in (22), and ${\cal L}'_2$
changes as
$${\cal L}'_2 \to \bar{\cal L}'_2 = {\cal L}'_2 + \Lambda \
d\left[ \theta^a \epsilon_{ab} e^b + {1\over 2} \theta^2\omega - {1\over 2}
d\theta^a \epsilon_{ab} \theta^b \right] \eqno(25)$$
Upon integration the boundary contributions involving $e^a$ and
$\omega$
may be dropped with the hypothesis that the dynamical variables
vanish on the boundary.  But because gauge parameters need not vanish, the
last term can survive, even though it is a total derivative.
$$I'_2 \to \bar{I}'_2 = I'_2 - \Lambda \int d^2x \left( \det
\partial\theta^a/\partial x^\mu\right) \eqno(26)$$

All the awkward features of the above formulation disappear if the gauge
theory is based on a {\it central extension\/} of the Poincar\'e algebra.  We
therefore postulate instead of (3), (14)
$$\left[ P_a,J\right] = \epsilon_a{}^b P_b \ \ ,\qquad \left[ P_a, P_b\right] =
\epsilon_{ab} {i\over 2}\Lambda I \eqno(27)$$
thereby adding the central element
$I$ to the generators and effecting a magnetic-like
modification of the translation algebra.$^9$  Consequently the connection
and curvature now become,
$$\eqalign{A &= e^a P_a + \omega J + a {i\over 2}\Lambda  I \cr
F &= dA + A^2 = f^a P_a + fJ + g{i\over 2}\Lambda
I = (De)^a P_a + d\omega J + \left( da +
{1\over 2}e^a \epsilon_{ab} e^b\right) {i\over 2}\Lambda I \cr}\eqno(28)$$
and the finite gauge transformations with gauge generator
$$\Theta = \theta^a P_a + \alpha J + \beta {i\over 2}\Lambda I  \eqno(29)$$
read
$$\eqalign{e^a \to \bar{e}^a &= \left( {\cal M}^{-1}\right)^a_{\ b} \left( e^b
+ \epsilon^b{}_c \theta^c\omega + d\theta^b\right) \cr
\omega \to \bar{\omega} &= \omega + d\alpha \cr
a \to \bar{a} &= a - \theta^a \epsilon_{ab} e^b - {1\over 2} \theta^2\omega +
d\beta + {1\over 2} d\theta^a \epsilon_{ab} \theta^b \cr}\eqno(30)$$
The multiplet of curvatures transforms by the adjoint representation of the
extended group,
$$\eqalign{f^a \to \bar{f}^a &= \left( {\cal M}^{-1}\right)^a_{\ b} \left( f^b
+ \epsilon^b{}_c \theta^c f\right) \cr
\bar{f}\to \bar{f} &= f \cr
g \to \bar{g} &= g - \theta^a \epsilon_{ab} f^b - {1\over 2}\theta^2 f
\cr}\eqno(\hbox{31a})$$
or
$$\eqalign{ F^A\to \bar{F}^A &= \sum^3_{B=0} \left( U^{-1}\right)^A_{\ B} F^B
\cr
U&= \left( \matrix{ {\cal M}^a{}_b & - \epsilon^a{}_c \theta^c & 0
\cr\noalign{\vskip 0.2cm}
0 & 1 & 0 \cr\noalign{\vskip 0.2cm}
\theta^c\epsilon_{cd} {\cal M}^d{}_b & - \theta^2/2 & 1 \cr}\right)
\cr}\eqno(\hbox{31b})$$
Note that in the above realization of the gauge action on $F$,
the extension is not
visible; $I$ is represented by ${\bf O}$.

An invariant Lagrange density is simply constructed with an extended multiplet
of Lagrange multipliers,
$$\eqalign{{\cal L}''_2 &= \sum^3_{A=0} \eta_A F^A = \eta^a (De)^a + \eta_2
d\omega + \eta_3 \left( da+{1\over 2} e^a\epsilon_{ab} e^b\right) \cr
\eta_A &= \left( \eta_a, \eta_2, \eta_3\right) \cr}\eqno(32)$$
which obey the conventional coadjoint transformation law,
$$\eqalignno{\eta_A \to \bar{\eta}_A &= \sum^3_{B=0} \eta_B
U^B{}_A &(\hbox{33a})\cr\noalign{\hbox{or}}
\eta_a \to \bar{\eta}_a &= \left( \eta_b - \eta_3 \epsilon_{bc}
\theta^c\right) {\cal M}^b{}_a\cr
\eta_2 \to \bar{\eta}_2 &= \eta_2  - \eta_a \epsilon^a{}_b \theta^b - {1\over
2}  \eta_3 \theta^2 &(33\hbox{b})\cr
\eta_3 \to \bar{\eta}_3 &= \eta_3 \cr}$$
Verlinde's
affine transformations for $\eta_a,\eta_2$ are now linear in $\eta_3$, which
is invariant.

Of course the equations of motion for ${\cal L}''_2$ are equivalent to those
for ${\cal L}'_2$, because variation of $a$ in (32) gives
$$d\eta_3 = 0 \eqno(34)$$
Therefore $\eta_3$ is constant, set equal to $\Lambda$, in which case ${\cal
L}''_2$ differs from ${\cal L}'_2$ by the total derivative
$\Lambda da$; but it is the
presence of this term that renders ${\cal L}''_2$ invariant in contrast to
${\cal L}'_2$.  (Clearly the solution $\eta_3=0$ gives an unextended
Poincar\'e gauge theory with vanishing cosmological constant.) Note that
varying the additional Lagrange multiplier $\eta_3$ provides one more equation
of motion
$$da + {1\over 2} e^a \epsilon_{ab} e^b = 0\eqno(35)$$
which can be always solved, at least locally, because the second term is a
closed 2-form.

The extended algebra, in the representation that we are using,
 possess a non-singular Killing metric, $h_{AB} = \sum_{CD} U^C{}_A h_{CD}
U^D{}_B$,
which is unavailable
without the extension.
$$h_{AB} = \left( \matrix{
h_{ab} &
\phantom{-}0 & \phantom{-}0 \cr\noalign{\vskip 0.2cm}
\phantom{-}0 & \phantom{-}0 & -1 \cr\noalign{\vskip 0.2cm}
\phantom{-}0 & -1 & \phantom{-}0 \cr}\right) \eqno(36)$$
This allows constructing the invariant scalar
$$M = -{1\over 2\Lambda} \sum^3_{B=0} \eta_A h^{AB} \eta_B\eqno(37)$$
which is also constant by virtue of the equations of motion and is interpreted
as the black hole mass.$^4$

In conclusion we see that the class of lineal gravity theories involving a
non-geometric Lagrange multiplier possesses two members that are distinguished
from the perspective of gauge invariance: the original model (1), (2) based on
the $SO(2,1)$ group and the string-inspired model (10) based on
the extended Poincar\'e group.  The
extended Poincar\'e
model involves an unconventional contraction of the $SO(2,1)$ model: owing to
the well-known ambiguity of two-dimensional angular momentum,
in (3)  one may replace $J$ by $J+sI/i$, $\Lambda$ by $\Lambda/s$ and set $s$
to infinity, thereby arriving at (27).  Finally we recall
that the de~Sitter model (2), (8) can be obtained by
 dimensional reduction of planar gravity; whether
there exists a model in $(2+1)$ dimensions that reduces to the
string-inspired lineal theory (10), (32) is under investigation.
\goodbreak
\bigskip
\centerline{\bf ACKNOWLEDGEMENT}
\medskip
\centerline{We thank S. Deser for helpful comments.}
\vfill
\eject
\centerline{\bf REFERENCES}
\medskip
\item{1.}C. Teitelboim, {\it Phys. Lett.\/} {\bf B126}, 41 (1983) and in {\it
Quantum Theory of Gravity\/}, S. Christensen, ed. (Adam Hilger, Bristol,
1984); R. Jackiw, in {\it Quantum Theory of Gravity\/},
S. Christensen, ed. (Adam Hilger, Bristol, 1984) and {\it Nucl. Phys.\/} {\bf
B252}, 343 (1985).
\medskip
\item{2.}For reviews see R. Mann, in {\it Proceedings of the Fourth Canadian
Conference on General Relativity and Relativistic Astrophysics\/}, to be
published; as well as J. Russo and A. Tseytlin, preprint.
\medskip
\item{3.}K. Isler and C. Trugenberger, {\it Phys. Rev. Lett.\/} {\bf 63}, 834
(1989); A. Chamseddine and D. Wyler, {\it Phys. Lett.\/} {\bf B228}, 75
(1989).
\medskip
\item{4.}H. Verlinde, in {\it Proceedings of the Sixth Marcel Grossmann
Meeting\/}, to be published.
\medskip
\item{5.}C. Callan, S. Giddings, A. Harvey and A. Strominger, {\it Phys.
Rev. D\/} {\bf 45}, 1005 (1992).
\medskip
\item{6.}Expression (10) is related to the usual
formula
$$I_2 = \int d^2x\,\sqrt{-g}\, e^{-2\varphi}\left( R - 4 \partial_\mu
\varphi \partial^\mu \varphi - \Lambda\right)$$
by rescaling the metric with the ``dilaton'' field $\varphi:g_{\mu\nu}\to
e^{2\varphi} g_{\mu\nu}$ and defining $\eta = e^{-2\varphi}$.
\medskip
\item{7.}S. Elitzur, A. Forge and E. Rabinovici, {\it Nucl. Phys.\/} {\bf
B359}, 581 (1991); G. Mandal, A. Sengupta and S. Wadia, {\it Mod. Phys.
Lett.\/} {\bf A6}, 1685 (1991); E. Witten, {\it Phys. Rev. D\/} {\bf 44}, 314
(1991).
\medskip
\item{8.}J. Russo, L. Susskind and L. Thorlacius, preprint; T. Banks, A.
Dabhokar, M. Douglas and M. O'Loughlin, preprint; S.~Hawking, preprint.
\medskip
\item{9.}This corresponds to a two-cocycle in the Poincar\'e group composition
law.
$$G\left({\cal M}_1, q_1\right) G\left({\cal M}_2,q_2\right)
= e^{{i\over 4} \Lambda q^a_1
\epsilon_{ab} \left({\cal M}_1q_2\right)^b }
G\left({\cal M}_1 {\cal M}_2,q_1+{\cal M}_1 q_2\right)$$
\par
\vfill
\end